# Data Accuracy Model for Distributed Clustering Algorithm based on Spatial Data Correlation in Wireless Sensor Networks

[1]Jyotirmoy Karjee ,[2]H.S Jamadagni

[1]Centre for Electronics Design and Technology, Indian Institute of Science, Bangalore, India
kjyotirmoy@cedt.iisc.ernet.in
[2]Centre for Electronics Design and Technology, Indian Institute of Science, Bangalore, India
hsjam@cedt.iisc.ernet.in

***Abstract***

*Objective: The main objective of this paper is to construct a distributed clustering algorithm based upon spatial data correlation among sensor nodes and perform data accuracy for each distributed cluster at their respective cluster head node. Design Procedure/Approach: We investigate that due to deployment of high density of sensor nodes in the sensor field, spatial data are highly correlated among sensor nodes in spatial domain. Based on high data correlation among sensor nodes, we propose a non - overlapping irregular distributed clustering algorithm with different sizes to collect most accurate or precise data at the cluster head node for each respective distributed cluster. To collect the most accurate data at the cluster head node for each distributed cluster in sensor field, we propose a Data accuracy model and compare the results with Information accuracy model. Finding: Simulation results shows that our propose Data accuracy model collects more accurate data and gives better performance than Information accuracy model at the cluster head node for each respective distributed cluster in our propose distributed clustering algorithm.Morover there exist a optimal cluster of sensor nodes which is adequate to perform approximately the same data accuracy achieve by a cluster. Practical Implementation: Measuring humidity and moisture content in an agricultural field, measuring temperature in physical environment. Inventive /Novel Idea: A distributed clustering algorithm is proposed based on spatial data correlation among sensor nodes with Data accuracy model.*

## 1. Introduction

**R**ecent development of wireless technology and embedded system made a drastic improvement over wireless sensor networks. Due to ease of deployment and reliable cost, sensor networks are used in many applications to sense or collect the physical phenomenon of raw data for any event such as temperature, humidity, seismic event, fire, etc from the physical environment [1]. A small processing unit device called node captures the physical phenomenon of raw data from the physical environment. These nodes can process the raw data, communicate wirelessly among other nodes and finally transmits the collected raw data to the base station or sink node.

Generally the physically sensed data collected by the sensor nodes are spatially correlated [2] in the sensor field. If the deployed density of sensor nodes increases, the spatially proximal sensor observations are highly correlated [3] in the sensor field. Since the sensor observations are highly correlated among sensor nodes, the sensor nodes form distributed clusters [4] in the sensor field to minimize data collection cost [5]. According to literature survey, LEACH [6] demonstrates a clear concept about distributed dynamic cluster formation according to priori probability. Each distributed cluster has respective Cluster Head (CH) [7] node which aggregates the data collected from all the sensor nodes in the cluster and finally transmits the processed data to the sink node. Moreover SEP [8] gives the cluster formation in the heterogeneous sensor networks. Literature [3, 9] shows the spatial correlation of observed data among sensor nodes to form distributed clusters. A grid based clustering method proposed in literature [10] shows a spatial correlation model for cluster formation. Basically this type of theoretical clustering model rarely happens in practical scenario in the sensor field. A disk-shaped circular cluster proposed in literature [11] shows grouping of nodes into disjoint set each managed by a designated CH node. However formation of disk shape cluster doesn't really appears in original scenario. Most of the cases, cluster formation are irregular in shape and size the in spatial domain. In literature [4] authors proposed a distributed clustering algorithm with different shape and size based upon shortest distance among sensor nodes and CH nodes in spatial domain. Here in this paper, we propose a formation of distributed clustering algorithm based upon spatially correlated data among sensor nodes. Our propose model for distributed clustering algorithm which form irregular shape and size is much more practical than the previously proposed clustering algorithm in spatial domain. As the numbers of sensor nodes are more in the sensor field, the data correlation among the sensor nodes increases [3] and form distributed clusters for high density of sensor nodes in our clustering algorithm. Thus finally we form spatially correlated distributed irregular non overlapping clusters of different sizes with high density of sensor nodes in spatial domain. More over the size of each distributed cluster in our algorithm is based upon a threshold value given in data correlation model [4] in spatial domain.

In literature [12, 13, 14], authors proposed Information accuracy (distortion function) model where base station or sink node can estimate the information accuracy for observed data sensed by all the sensor nodes. These types of model are based on one hop communication where observed data are sensed by all the sensor nodes and directly transmit the observed data to the sink node. But in literature [4, 15] authors proposed two hop communication where observed data are transmitted to the sink node via intermediate node (CH node) where the sensor field is large. Again in this paper, we consider two hop communications for our distributed clustering algorithm based on spatial data correlation among sensor nodes in which observed data are transmitted to the sink node via CH node. From literature survey, it has been noted that estimated data collected from all the sensor nodes in a cluster are directly send to CH node for aggregation[24,25] without verifying the accuracy. Hence it is important to verify the estimated data before data aggregation at CH node and then send it to the sink node. For each distributed cluster, the data accuracy is verified using MMSE estimator [23] before data aggregation and then only transmits the most accurate data to the sink node. Thus verifying data accuracy at CH node before data aggregation for each distributed cluster may reduce communication overhead. It may possible that some of the sensor nodes in the distributed cluster get malicious [16] due to external physical environment .In such tropical situation sensor nodes can sense and read inaccurate data. These inaccurate data transmitted by malicious nodes may cause incorrect data aggregation at the CH node for respective clusters. Hence it is required to estimate and verify the data accuracy before data aggregation in the CH node for each distributed cluster to reduce data redundancy and power consumption.

In this paper, we propose Data accuracy model where we use Minimum Mean Square Error (MMSE) estimation to perform data accuracy at the CH node before data aggregation [17] for each distributed cluster. Most of the work done [12, 18] till today is to perform MMSE estimation at each individual sensor nodes for the observed data before transmitting the estimated data at the CH[1] node in a cluster. According to literature [18], once the estimated data is received at the CH node transmitted by all the sensor nodes in a cluster, averaging the estimated data at CH node and finally transmits the most accurate data to the sink node. However to the best understanding of authors, this is the first time to perform MMSE estimation only at the CH node for all the observed data sensed by all the sensor nodes in a cluster .In our Data accuracy model, calculating MMSE estimation only at the CH node for the observed data sensed by all the sensor nodes in a cluster can increase the data accuracy and reduce the communication overhead before data aggregation.

Rest of the paper is given as follows. In section-2, we construct a data correlation model [4] among sensor nodes in spatial domain. Data correlation model shows the degree of correlation coefficient for observed data among sensor nodes. The degree of correlation coefficient for observed data are measured by an assumed threshold value. If the correlation coefficients for observed data among sensor nodes are greater than the threshold value, observed data are spatially correlated among sensor nodes in spatial domain otherwise not. Ultimately from this threshold value, we get an approximated circular data correlation range among sensor nodes. The size of this approximated circular data correlation range depends upon the threshold value. The sensor nodes which fall with in this circular data correlation range, the spatial data among them are highly correlated in the spatial domain. Hence the correlation coefficients for observed data among these sensor nodes are greater than the threshold value. In section-3, we propose a distributed clustering algorithm based upon spatial correlation for observed data among sensor nodes in the sensor field .It forms non over-lapping irregular shape and size of different distributed clusters in the spatial domain. Once the distributed clusters are formed in the sensor field, each cluster can perform the data accuracy at their respective CH node and transmit the most accurate data to the sink node which is discussed in section-4. We also construct a Data accuracy model and compare it with Information accuracy model with respect to data accuracy. In section 5, we perform the simulation and validation for our proposed distributed clustering algorithm and Data accuracy model. Finally we conclude our work in section 6.

## 2. Data Correlation Model in Spatial Domain

In this section, we are interested to illustrate the spatial data correlation among sensor nodes $i$ and $j$ to sense or measure a tracing point [4] $s^i$ for $i = 1$ in a spatial domain. Tracing point is a reference value which we are interested to measure and sense in the spatial domain. For example tracing point has higher concentration of moisture content in an agricultural field. It has higher concentration of data with higher variation with respect to lower variation of data in the spatial domain. As the sensor node density increases, the spatial correlation of observed data $(S_i, S_j)$ among the sensor nodes $i$ and $j$ also increases in the spatial domain. The sensor nodes sense and measure the tracing point over a window frame of time interval $T$ to capture the continuous data sample with $S_i=\{ s_{i1} ,\ s_{i2}, s_{i3}, \ldots\ldots s_{in} \}$ and $S_j=\{s_{j1} ,\ s_{j2}, s_{j3}, \ldots\ldots s_{jn}\}$ respectively. If the tracing point sensed and measured by the sensor nodes $i$ and $j$ located near to each other, the data correlation is strong. The data correlation decreases as the sensor nodes $i$ and $j$ are far apart from the tracing point. The sensor nodes $i$ and $j$ can compute the mean of continuous data sample over a window frame of time interval *T*. Thus the mean of continuous data sample sensed and measured by sensor nodes $i$ and $j$ are given as follows.

[1] According to literature [18] CH node is only a logical entity and can also be called as sink node depending upon applications.

$$\bar{S}_i = \frac{1}{n}\sum_{k=1}^{n} S_{ik} \quad \text{and} \quad \bar{S}_j = \frac{1}{n}\sum_{k=1}^{n} S_{jk} \tag{1}$$

We compute the variance of continuous sample data captured by the sensor nodes $i$ and $j$ in spatial domain. Variance is used to measure how far a set of continuous data sample of sensor nodes are spread out from each other. Thus the sensor nodes compute the variance of sample data as follows.

$$Var(S_i) = \frac{1}{n-1}\sum_{k=1}^{n}(S_{ik} - \bar{S}_i)^2 \quad \text{and} \quad Var(S_j) = \frac{1}{n-1}\sum_{k=1}^{n}(S_{jk} - \bar{S}_j)^2 \tag{2}$$

We compute the covariance of sample data for nodes $i$ and $j$ which is given as

$$Cov(S_i, S_j) = \frac{1}{(n-1)}\sum_{k=1}^{n}(S_{ik} - \bar{S}_i)(S_{jk} - \bar{S}_j) \tag{3}$$

Covariance is defined as a measure of how much two variable of continuous sample data change together in a spatial domain for sensor nodes $i$ and $j$ .We find the correlation coefficient ( $\rho_{S_iS_j}$ ) for spatial correlation between sample data $(S_i, S_j)$ sensed by the sensor nodes $i$ and $j$ which is given as

$$\rho_{S_iS_j} = \frac{Cov(S_iS_j)}{Var(S_i)Var(S_j)}$$

$$\rho_{S_iS_j} = \frac{\frac{1}{(n-1)}\sum_{k=1}^{n}(S_{ik} - \bar{S}_i)(S_{jk} - \bar{S}_j)}{\left[\frac{1}{(n-1)}\sum_{k=1}^{n}(S_{ik} - \bar{S}_i)^2\right]\left[\frac{1}{(n-1)}\sum_{k=1}^{n}(S_{jk} - \bar{S}_j)^2\right]} \tag{4}$$

Thus equation no. (4) shows the data correlation between the sample data among sensor nodes $i$ and $j$ in the spatial domain. These spatially correlated data among sensor nodes $i$ and $j$ can be modeled as Joint Gaussian Random Variables (JGRV) [12, 14] as follows:

$E[S_i] = 0$ , $E[S_j] = 0$ *for i=1,2,…………..n and j=1,2 ……..n*

$Var[S_i] = \sigma_{Si}^2$ , $Var[S_j] = \sigma_{Sj}^2$ *for i=1,2………n and j=1,2,……..n*

$Cov[S_i, S_j] = \sigma_{Si}^2 Corr[S_i, S_j]$

$\sigma_{Si}^2 Corr[S_iS_j] = \sigma_{Si}^2 \rho[S_iS_j] = E[S_iS_j] = Cov[S_iS_j]$

$$K_V(d_{i,j}) = Corr[S_i S_j] = \rho[S_i S_j] = \frac{E[S_i S_j]}{\sigma^2_{S^i}} = \frac{Cov[S_i S_j]}{\sigma^2_{S^i}} \quad (5)$$

$K_V(.)$ is a correlation model [14]and the Euclidian distance between the sensor nodes $i$ and $j$ can be represented as $d_{i,j} = \| S_i - S_j \|$ for the sensed data . We assume the covariance function to be non-negative and can decrease monotonically with distance $d_{i,j} = \| S_i - S_j \|$,with limiting value of 1 at $d = 0$ and of 0 at $d = \infty$ . We adopt power exponential model [19, 20] which is given as

$$K_V^{PE}(d_{i,j}) = e^{(-d/\theta_1)^{\theta_2}} \quad \text{for } \theta_1 > 0 \ , \ \theta_2 > (0,2] \quad (6\ )$$

where $\theta_1$ is called range parameter which controls the relation between the distance among sensor nodes and the correlation coefficient. It also controls how fast the correlation decays with distance among the sensor nodes. $\theta_2$ is called a smoothness or roughness parameter which controls geometrical properties of the random field. It contains exponential model for $\theta_2 = 1$ and squared exponential model for $\theta_2 = 2$ .From equations no. (5) and (6), we find the correlation coefficient of observed data $S_i(x_i, y_i)$ as well as $S_j(x_j, y_j)$ among the sensor nodes $i$ and $j$ using power exponential model as follows

$$\rho[S_i S_j] = e^{-\left(\frac{d_{i,j}}{\theta_1}\right)^{\theta_2}} \quad (7)$$

We define a threshold $\tau$ for $0 < \tau \le 1$ which determines whether the spatial data are correlated among the sensor nodes in the sensor field. Using the threshold value $\tau$ , we show two properties for spatially correlated data among sensor nodes as follows:

- If $\rho[S_i S_j] \ge \tau$ , spatial data are strongly correlated among sensor nodes $i$ and $j$ in the spatial domain.
- If $\rho[S_i S_j] < \tau$ , spatial data are weakly correlated among sensor nodes $i$ and $j$ in the spatial domain.

From equations no. (4 ), (6 ) and (7 ), we define the correlation coefficient $\rho[S_i S_j]$ for the observed data using power exponential model among sensor nodes $i$ and $j$ where the data are strongly correlated in the spatial domain represented as follows

$$\rho[S_i S_j] = \frac{Cov[S_i S_j]}{\sigma^2_{S_i}} = e^{-\left(\frac{d_{i,j}}{\theta_1}\right)^{\theta_2}} \ge \tau \quad (8)$$

From equation no 8, we find the relation between the threshold values $\tau$ and power exponential model represented as

$$e^{-\left(\frac{d_{ij}}{\theta_1}\right)^{\theta_2}} \geq \tau$$

$$d_{ij}^2 \leq \theta_1^2 \sqrt[\theta_2]{\left(\log\left(\frac{1}{\tau}\right)\right)^2} \tag{9}$$

We compare the equation no (9) with the Euclidean distance among the coordinates of sensor nodes $i$ and $j$ as follows.

$$d_{ij}^2 = (x_i - x_j)^2 + (y_i - y_j)^2 \tag{10}$$

From equations no (9) and (10), we get

$$(x_i - x_j)^2 + (y_i - y_j)^2 \leq \theta_1^2 \sqrt[\theta_2]{\left(\log\left(\frac{1}{\tau}\right)\right)^2} \tag{11}$$

Comparing equation no. (11) with the equation of a circle , we get

$$(x_i - x_j)^2 + (y_i - y_j)^2 = r^2 \tag{12}$$

From equations no. *(11) and (12 )*, we find the radius $r$ for range of circular data correlation area denoted as $cir(i)$ around a sensor node $i$ as a centre coordinate .

$$r^2 \leq \theta_1^2 \sqrt[\theta_2]{\left(\log\left(\frac{1}{\tau}\right)\right)^2} \tag{13}$$

The sensor nodes $j$ which falls under $cir(i)$, the observed data among sensor nodes $i$ and $j$ are highly correlated in the spatial domain. The spatial data correlation $\rho[S_i S_j]$ among sensor nodes $i$ and $j$ with in $cir(i)$ are greater than the threshold value $\tau$ .

Equation no. (13), shows that the radius $r$ of circular data correlation area $cir(i)$ depends upon the threshold value $\tau$ , $\theta_1$ and $\theta_2$ in the spatial domain. We define two properties from equation no. (13) given as follows:

- For a fixed value of $\theta_1$ and $\theta_2$ , if the threshold $\tau$ increases, the radius $r$ of $cir(i)$ decreases exponentially.
- Similarly with a fixed value of $\theta_1$ and $\theta_2$, if the radius $r$ of $cir(i)$ increases, the size of $cir(i)$ also get increase and the average number of distributed clusters (discuss in section-5) decreases exponentially in the sensor region.

Hence we take an appropriate threshold value $\tau$ to find the size of $cir(i)$ where the observed data among sensor nodes are strongly correlated in the spatial domain. In the next section, we propose a distributed clustering algorithm based upon data correlation among sensor nodes in each $cir(i)$ in the sensor field.

## 3. Distributed Clustering Algorithm based on Spatial Data Correlation

In this section, we propose a distributed clustering algorithm which forms non overlapping clusters of irregular shape and size in the sensor field. If the deployed sensor nodes increases in the sensor field, the spatial data correlation among the sensor nodes increases. Based upon the spatial data correlation among the sensor nodes for each $cir(i)$ in the sensor field, we construct the distributed clustering algorithm.

*Notations used in the clustering algorithm:*

$M$ = Total number of sensor nodes in deployed sensor field
$i$ = Represents each sensor node where $i \in M$
$id(i)$ =Represents identification number of each sensor node $i$
$cir(i)$ =Range of data correlation area which is approximated by a circular area around each sensor node $i$ as centre
$r$ =Radius of data correlation range area $cir(i)$
$G(i)$ =A group of neighboring sensor nodes $j$ which is a subset of $cir(i)$ of node $i$ as centre
$\max NodeG(i)$ =Maximum number of sensor nodes $j$ in $G(i)$ of $cir(i)$
$\max DisG(i)$ =Maximum Euclidian distance between the farthest node $j$ from node $i$ as a centre in $\max NodeG(i)$
$\min SizeG(i)$ =Minimum size of $\max DisG(i)$
$\rho[S_i S_j]$ =Spatial data correlation coefficient between nodes $i$ and $j$
$\tau$ = Threshold value
$W$ = Set of $id(i)$ which doesn't form cluster
$\bar{W}$ = Set of $id(i)$ which form cluster

**Distributed Clustering Algorithm**

*Step 1:* Start
*Step 2:* Initially $W = \{M\}$ where $id(i) \in M$ for $i = 1, 2, ..............M$ and $\bar{W} = \{\varnothing\}$
*Step 3:* For each $i$, $G(i) = \{j : d(i, j) \le r, i \ne j\}$ where $d(i, j)$ is the Euclidian distance between $i$ and $j$
*Step 4:* if $G(i) \subseteq cir(i)$ ,then $\rho[S_i S_j]$ is strongly correlated and $\rho[S_i S_j] \ge \tau$
*Step 5:* Compute $G(i)$ for $cir(i)$ of each sensor node $i$
*Step 6:* Check for $\max NodeG(i)$ in each $cir(i)$ in the sensor field

If more than one same $\max NodeG(i)$ in the sensor field
{
Compute $\max DisG(i)$ for all $\max NodeG(i)$
Compute $\min SizeG(i)$ among $\max DisG(i)$
$\min SizeG(i)$ form the cluster among $\max NodeG(i)$ with CH node as $i$ and add each $id(i)$ in $\bar{W}$
}

else

$\max NodeG(i)$ form the cluster with CH node $i$ and add each $id(i)$ in $\bar{W}$

*Step 7:* Repeat *Step 6* until $W = \{\varnothing\}$ and $\bar{W} = \{M\}$ where each $id(i) \in M$ for all $i = 1, 2..........M$ where $G(i) \cap G(j) = \varnothing$ in the sensor field

*Step 8:* Stop

---

We consider a rectangular sensor field where $M$ sensor nodes are randomly deployed. We assume that every sensor nodes knows the position of the coordinates of all sensor nodes in the sensor field like MTE routing [6] to simplify the deployment topology. In the previous section, we clarify that for a threshold value $\tau$, we get the radius $r$ of $cir(i)$ for each sensor node $i$ in the sensor field. Hence we fix a threshold value $\tau$ for which we get radius $r$ of an appropriate size of $cir(i)$ for each sensor node $i$ in the sensor region. This means that each sensor node $i$ perform the data correlation with the neighboring [26, 27] sensor nodes $j$ to form $G(i)$ within the data correlation range area $cir(i)$. $cir(i)$ is approximated by a circular area around the node $i$ with radius of data correlation range $r$. $G(i)$ includes the node $i$ itself as the centre of $cir(i)$ and the neighboring nodes $j$ which fall under the data correlation range of area $cir(i)$ with radius $r$. Thus $G(i)$ for data correlation among the sensor nodes $i$ and $j$ with in $cir(i)$ can be given as

$$G(i) = \{ j : d(i, j) \leq r, i \neq j\} \tag{14}$$

Where $d(i, j)$ is the Euclidian distance between sensor nodes $i$ and $j$. The spatial data correlation for $G(i)$ of $cir(i)$ are partially or fully overlapped with $G(j)$ of $cir(j)$ in the sensor field. Thus overlapping of many data correlation range area occurs in the sensor field. Overlapping of spatial data correlation for $cir(i)$ and $cir(j)$ can share the same correlated overlapping of data among $G(i)$ and $G(j)$. Thus overlapping of same correlated data is like utilizing the same resource [4] among the $G(i)$ and $G(j)$ in the data correlation range areas $cir(i)$ and $cir(j)$. Hence it increases the data redundancy among $cir(i)$ and $cir(j)$. Hence a distributed clustering algorithm is proposed to overcome the overlapping problem of spatially correlated data among $G(i)$ and $G(j)$. Thus the distributed clustering formation consists of the following phases:

*Phase-I:*

Each sensor node $i$ has its node identification number $id(i)$ for $W = \{M\}$ where $id(i) \in M$, $i = 1, 2, ..............M$ and $\bar{W} = \{\varnothing\}$. For each sensor node $i$, $id(i) \in M$ which participate to form cluster in later phase, leaves from the array $W$ and add to an array $\bar{W}$. $\bar{W}$ is an array which signifies that each sensor node $i$ of $id(i)$ participate to form cluster.

*Phase-II:*

Each sensor node $i$ computes $G(i)$ with in the data correlation range area $cir(i)$ with radius $r$ and satisfies the equation no. (14).

*Phase-III:*
Check for each sensor node $i$ having $\max NodeG(i)$ of $cir(i)$ in the sensor field which forms the first cluster in the sensor region. Sensor node $i$ form the CH node of $\max NodeG(i)$ in $cir(i)$ .Hence $\max NodeG(i)$ forms the cluster in $cir(i)$ leaves from the array $W$ and added to array $\bar{W}$ .

*Phase-IV:*
If there are more than one same $\max NodeG(i)$ of $cir(i)$ in the sensor field, then there is a big question that which $\max NodeG(i)$ forms the cluster. This problem can be resolve in two steps:

- Firstly we compute $\max DisG(i)$ for all $\max NodeG(i)$ of $cir(i)$ in the sensor field.
- Secondly we find the $\min SizeG(i)$ among $\max DisG(i)$ for all $\max NodeG(i)$ of $cir(i)$ in the sensor field. We calculate $\min SizeG(i)$ among $\max DisG(i)$ because the data correlation among closer nodes for $\min SizeG(i)$ are strong to form cluster.

Hence $\min SizeG(i)$ forms the cluster among $\max DisG(i)$ with sensor node $i$ as CH node and add each $id(i)$ in array $\bar{W}$ .

*Phase-V:*
Repeat *Phases-III* & *IV* until $W = \{\varnothing\}$ and $\bar{W} = \{M\}$ where $id(i) \in M$ for all $i = 1, 2..........M$ .
Finally all the sensor nodes participate to form non overlapping distributed clusters with $G(i) \cap G(j) = \varnothing$ in the sensor field.

Therefore we construct a non overlapping distributed clustering algorithm in this section based upon spatial data correlation among sensor nodes in the sensor field. In the next section we are interested to find the data accuracy estimation for each distributed cluster and send the most accurate data to the sink node.

## 4. Distributed Cluster-based Data Accuracy Model

In the previous section, we develop a non-overlapping distributed clustering algorithm with irregular shape and size in the sensor field based upon data correlation among sensor nodes. We assume each distributed cluster can sense and measure a single tracing point of same event and perform the data accuracy for the measured data at the CH node for the respective cluster. Finally CH node of each distributed cluster transmits the most accurate data to the sink node in the sensor region. Each distributed cluster has different set of sensor nodes to perform the data accuracy. The data accuracy is perform to verify the estimated data received at the CH node from all the sensor nodes for a cluster are most accurate and doesn't contain any redundant data in it. It may reduce the communication overhead.

For the simplest analysis of our propose Data accuracy model, we choose a single cluster of $M$ sensor nodes. Cluster with $M$ sensor nodes can sense a single tracing point and check the data accuracy at the CH node before data aggregation and then transmit the most accurate data to the sink node. Here we demonstrate the mathematical analysis of data accuracy for a single cluster with $M$ sensor nodes. Each sensor node $i$ can observe and measure the physically sensed data $S_i$ for the tracing point value $S$ with observation noise $n_i$ for the cluster. Therefore the observation done by the sensor node $i$ in a cluster is illustrated as

$$x_i = s_i + n_i \qquad \text{where } i \in M \tag{15}$$

The sensor node $i$ sense the observation sample data $x_i$ and transmits $x_i$ to the CH node sharing wireless Additive White Gaussian Noise(AWGN) channel [12 ,21 ] where $n_i$ is independent of each other and modeled as Gaussian Random Variable of zero mean and variance $\sigma_n^2$. Thus the observation sample data $x_i$ passes through AWGN channel to the CH node for the cluster which reconstructs estimation $\hat{S}$ of the tracing point $S$ . The CH node receive all $M$ observation sample for the cluster given by

$$X = AZ + N \tag{16}$$

where $X$ is a $M \times 1$ data vector for observation done by $M$ sensor nodes in a cluster , $Z$ is a $(1+M)\times 1$ random vector for physically sensed data $S_i$ for $i \in M$ including the point event $S$ where we estimate for $\mathcal{N}(0, C_Z)$ , A is a known $M \times (1+M)$ matrix and $N$ is a $M \times 1$ noise vector for the observed data of $M$ sensor nodes with $\mathcal{N}(0, C_Z)$.The random vector Z with zero mean and covariance $\mathcal{N}(0, C_Z)$ can be shown as follows :

$$\mu_z = E[Z] = \begin{bmatrix} E[s] \\ E[s_1] \\ E[s_2] \\ . \\ . \\ E[s_M] \end{bmatrix} = \begin{bmatrix} 0 \\ 0 \\ 0 \\ . \\ . \\ 0 \end{bmatrix} \quad \text{and} \quad C_z = E[ZZ^T] = \begin{bmatrix} E[s,s] & E[ss_1] & E[ss_2] & . & . & E[ss_M] \\ E[s_1 s] & E[s_1 s_1] & E[s_1 s_2] & . & . & E[s_1 s_M] \\ E[s_2 s] & E[s_2 s_1] & E[s_2 s_2] & . & . & E[s_2 s_M] \\ . & . & . & . & . & . \\ . & . & . & . & . & . \\ E[s_M s] & . & . & . & . & E[s_M s_M] \end{bmatrix}$$

Thus the covariance matrix is

$$C_z = \sigma_s^2 \begin{bmatrix} 1_{1\times 1} & R^T_{1\times M} \\ R_{M\times 1} & B_{M\times M} \end{bmatrix} \tag{17}$$

where

$$R_{M\times 1} = \begin{bmatrix} \rho_{s_1,s} \\ \rho_{s_2,s} \\ . \\ . \\ \rho_{s_M,s} \end{bmatrix} \qquad B_{M\times M} = \rho_{s_i,s_j} = \begin{bmatrix} \rho_{s_1,s_1} & \rho_{s_1,s_2} & . & \rho_{s_1,s_M} \\ \rho_{s_2,s_1} & \rho_{s_2,s_2} & . & \rho_{s_2,s_M} \\ . & . & . & . \\ \rho_{s_M,s_1} & \rho_{s_M,s_2} & . & \rho_{s_M,s_M} \end{bmatrix}$$

In $C_Z$ matrix, $R_{M\times 1} = \rho_{S_i,S}$ gives the correlation coefficient between $S_i$ , $S$ respectively and $B_{M\times M} = \rho_{S_i,S_j}$ gives the correlation coefficient between $S_i$ , $S_j$ respectively. Now the power exponential model [19, 20] can be used for correlation model to show the relation between $S_i$ and $S$ as well as $S_i$ and $S_j$. Thus we get $\rho_{S_i,S} = e^{-\left(d_{S,i}/\theta_1\right)^{\theta_2}}$ and $\rho_{S_iS_j} = e^{-\left(d_{S,i}/\theta_1\right)^{\theta_2}}$ in the covariance matrix $C_Z$ .

CH node collects all the observations from $M$ sensor nodes in the cluster to find the estimate of $\hat{S}$ from $\hat{S}_i$ . If the observed data $X$ can be modeled by Bayesian Linear Model [22 ] for all sensor nodes in cluster , the MMSE estimator to estimate the tracing point at the CH node in a cluster is given as :

$$\hat{Z} = E\left[Z \mid X\right] = \begin{bmatrix} \hat{s} \\ \hat{s}_1 \\ \hat{s}_2 \\ . \\ \hat{s}_M \end{bmatrix}$$

$$\hat{Z} = C_Z A^T (A C_Z A^T + \sigma_N^2 I_{M\times M})^{-1} X$$

$$\hat{Z} = \begin{pmatrix} R^T \\ B \end{pmatrix} \left( B + \frac{\sigma_N^2}{\sigma_S^2} I_{M\times M} \right)^{-1} X \tag{18}$$

The measurement of performance for the MMSE estimator at the CH node for the cluster is given as the error $\in = (S - \hat{S})$ with mean zero and covariance matrix illustrated as

$$E[(Z - \hat{Z})(Z - \hat{Z})^T]$$

$$= C_Z - C_Z A^T (A C_Z A^T + \sigma_N^2 I_{M\times M})^{-1} A C_Z$$

$$= \sigma_S^2 \begin{pmatrix} 1 & R^T \\ R & B \end{pmatrix} - \sigma_S^2 \begin{pmatrix} R^T \\ B \end{pmatrix} \left( B + \frac{\sigma_N^2}{\sigma_S^2} I_{M \times M} \right)^{-1} (\mathrm{R} \;\; B) \quad (19)$$

From equation no.(18), we get the estimation of tracing point $(\hat{S})$ at the CH node in a cluster given as

$$\hat{S} = R^T \left( B + \frac{\sigma_N^2}{\sigma_S^2} I_{M \times M} \right)^{-1} X \quad (20)$$

We find the distortion factor between $S$ and $\hat{S}$ to perform data accuracy at the CH node for a cluster. From equation no (19), we get the distortion factor as

$$\bar{D} = E[(S - \bar{S})^2]$$

$$\bar{D} = \sigma_S^2 - \sigma_S^2 R^T \left( B + \frac{\sigma_N^2}{\sigma_S^2} I_{M \times M} \right)^{-1} R \quad (21)$$

We normalize the distortion factor and calculated the data accuracy for $M$ sensor nodes for a cluster as

$$D_A(M) = 1 - \frac{\bar{D}}{\sigma_S^2}$$

$$D_A(M) = R^T \left( B + \frac{\sigma_N^2}{\sigma_S^2} I_{M \times M} \right)^{-1} R \qquad \text{where} \quad \left( B + \frac{\sigma_N^2}{\sigma_S^2} I_{M \times M} \right)^{-1} R = \beta$$

$$D_A(M) = R^T \beta \quad (22)$$

$D_A(M)$ calculated at the CH node for each distributed cluster is performed before data aggregation and finally send the most appropriate data to the sink node. Hence the purpose of verifying the data accuracy $D_A(M)$ at CH node for each distributed cluster is to confirm that the most accurate data transmitted by $M$ sensor node can aggregate rather than aggregating all the redundant data at the CH node. Once we perform the estimation to calculate the data accuracy $R^T \beta$ at the CH node for each distributed cluster, the most precise data get aggregated and finally send to the sink node.

The information accuracy model proposed in literature [18] shows that at first each sensor nodes $i$ can calculate the MMSE estimate $\hat{S}_i$ for observed data and then transmits the estimated data $\hat{S}_i$ to the CH node i.e. $\hat{S}_i$ in order to find $\hat{S}$. Finally averaging all $\hat{S}_i$ at the CH node for the cluster for $M$ sensor nodes to get $\hat{S}$.But in Data accuracy model $D_A(M)$ , at first we collect all the observed data from $M$ sensor nodes and then only perform the MMSE estimation at the CH node for each distributed cluster. It is better to perform the MMSE estimation only at the CH node rather than performing the MMSE estimation at individual nodes and then averaging it at the CH node for distributed cluster. We perform the MMSE at the CH node as it is the only central authority for each distributed cluster and it knows the activities of cluster members.

## 5. Simulation and Validation

Data correlation model discussed in section-2, shows that the spatial correlation for observed data ($S_i$ and $S_j$) among sensor nodes $i$ and $j$. Spatial correlations for observed data are strong when it is greater than some threshold value $\tau$. So we fixed a threshold value $\tau$ for $0 < \tau \leq 1$. Above this threshold value $\tau$ , spatial data are strongly correlated among sensor nodes $i$ and $j$. Depending upon the threshold value $\tau$ ,we get a radius $r$ for each sensor node $i$ to perform data correlation with neighboring node $j$,approximated by a circular data correlation range area $cir(i)$ around each node $i$. This means with in the range of $cir(i)$ for node $i$ with data correlation radius of $r$ , data are strongly correlated with other nodes $j$. So in the first simulation setup, we clarify the relation between $\tau$ and $r$. In the Fig. 1(a), we plot the relation between threshold values $\tau$ and the corresponding size of data correlation radius $r$ for node $i$ of $cir(i)$. If the threshold value $\tau$ increases for $0 < \tau \leq 1$ with $\theta_1$=70 and $\theta_2$=1, the radius $r$ of $cir(i)$ decreases exponentially. In Fig. 1(b), we show the sizes of data correlation radius $r$ for node $i$ of $cir(i)$ and the average number of clusters based on spatial correlation for threshold values $\tau$=0.5. If $r$ of $cir(i)$ increases, the size of $cir(i)$ increases for $\theta_1$=70, $\theta_2$=1 and the average number of distributed clusters based on spatial correlation (discussed in section -3) decreases exponentially.

In the second simulation set up, we have a sensor field of $2m \times 2m$ grid based sensor topology with CH node on one of the corner edge and a fixed tracing point located in the center as given in Fig .2 according to literature [18]. We deployed thirty four sensor nodes and a CH node which forms a cluster in grid based sensor topology. We are interested to demonstrate the data accuracy with respect to the number of sensor nodes. We set the same sensor field topology (Fig.2) as given in literature [18] where the position of sensing nodes are located at point (6,2),(8,4),(6,4),(4,4) and the tracing point at (6,4). For these four jointly sensing nodes, the information accuracy $I(M)$ in literature [18] is 0.7469 and our result for data accuracy $D_A(M)$ is 0.7545. This shows our propose Data accuracy model $D_A(M)$ give more accurate data than the Information accuracy model $I(M)$ proposed in literature [18] for the same sensor nodes with same topology. Moreover, if we introduce a fifth sensing node located at (10, 4), the information accuracy $I(M)$ is 0.7462. This clarifies that introduce of a fifth node which is far away from the tracing point dominates its observation results and decreases the information accuracy. But in our Data accuracy model $D_A(M)$,the introduce of fifth node may increase the data accuracy of 0.7665. Hence introduce of a new sensor nodes in the sensor field increase data accuracy in our propose $D_A(M)$. Fig.3 shows that the results for data accuracy $D_A(M)$ is always greater than information accuracy $I(M)$ as we keep increasing the number of sensor nodes for $\theta_1$=70 and $\theta_2$=1. Thus our propose Data accuracy model give more accurate data and better performance than the Information accuracy model with respect to number of sensor nodes in a cluster.

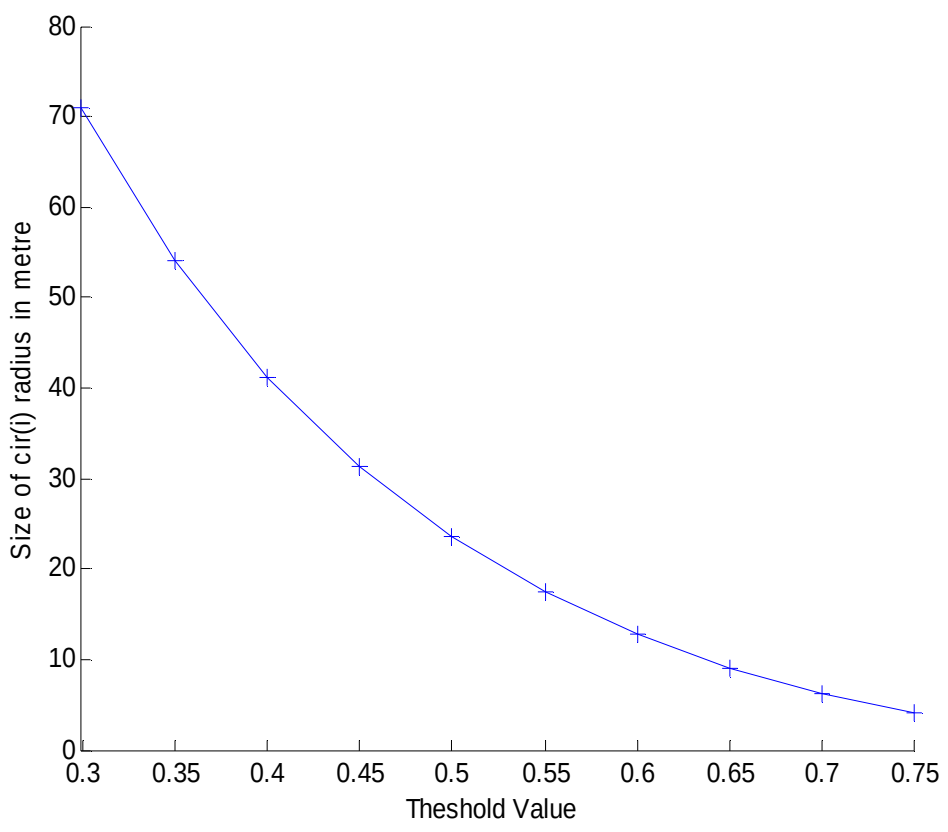

Fig.1 (a) Threshold value versus size of $cir(i)$ radius

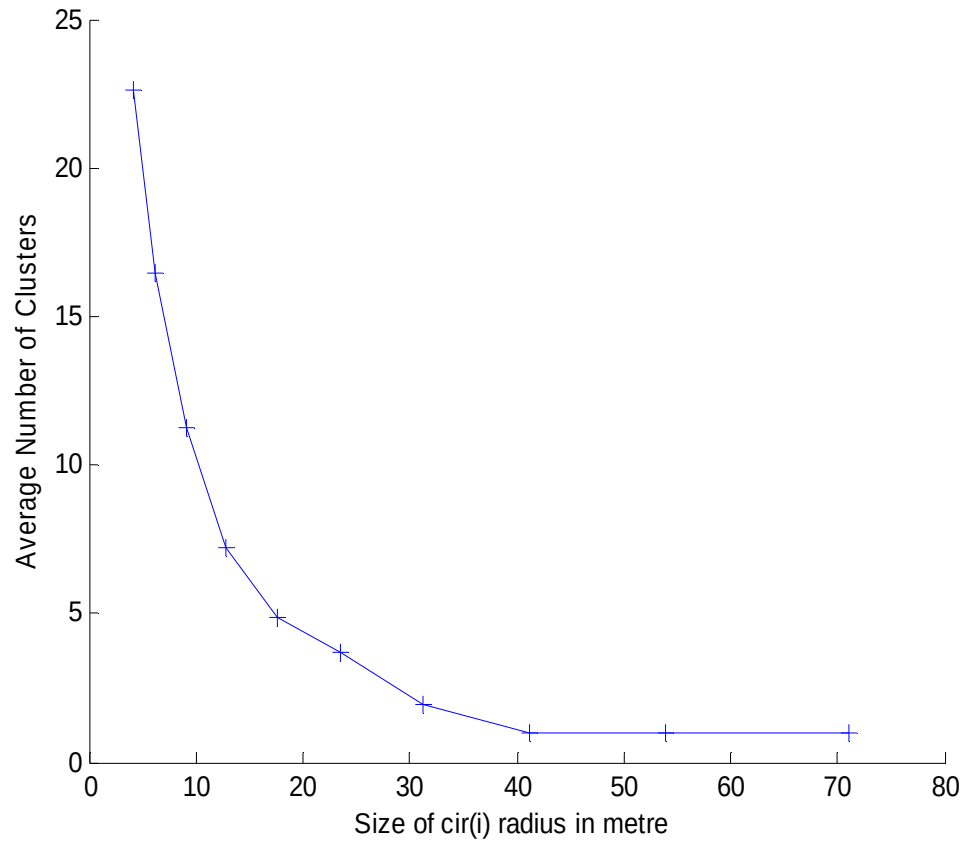

Fig.1(b) Size of $cir(i)$ radius versus average number of clusters

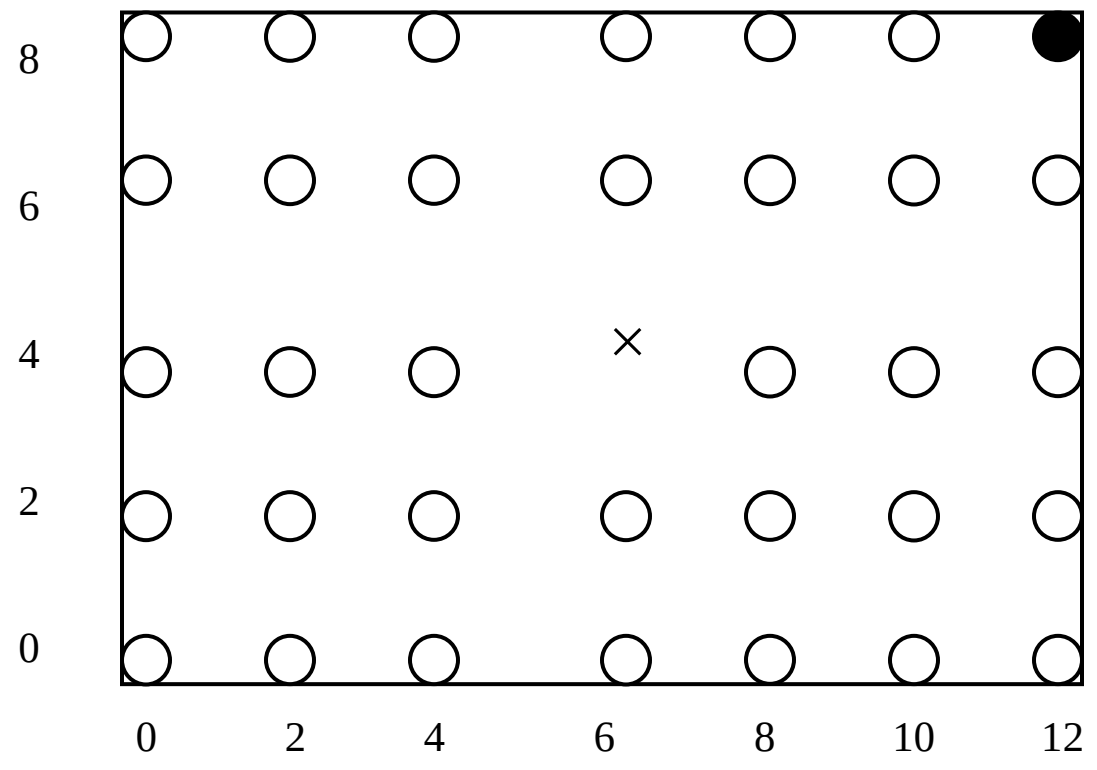

**Fig. 2** . Wireless sensor network topology: ○ means sensor node , ● means CH node , × means tracing point

Moreover, if we continuously increase the number of sensor nodes in the sensor field, the data accuracy remains approximately same. Fig. 3 shows fifteen to twenty sensor nodes are sufficient to perform the same data accuracy level which we achieve for thirty four sensor nodes. Hence we can reduce the number of sensor nodes in a cluster with respect to data accuracy. It is unnecessary to deploy thirty four sensor nodes beyond this upper bound because fifteen to twenty sensor nodes are sufficient to give approximately the same data accuracy level achieve in the cluster. Hence fifteen to twenty sensor nodes perform the data accuracy at the CH node for the cluster and transmit the accurate data to the sink node. Thus fifteen to twenty sensor nodes (optimal cluster) perform the communication process and rest of the sensor nodes goes to sleep mode in the cluster .Reducing the number of sensor nodes to fifteen to twenty sensor nodes instead of deploying thirty four sensor nodes can reduce communication overhead as well as energy consumption in a cluster.

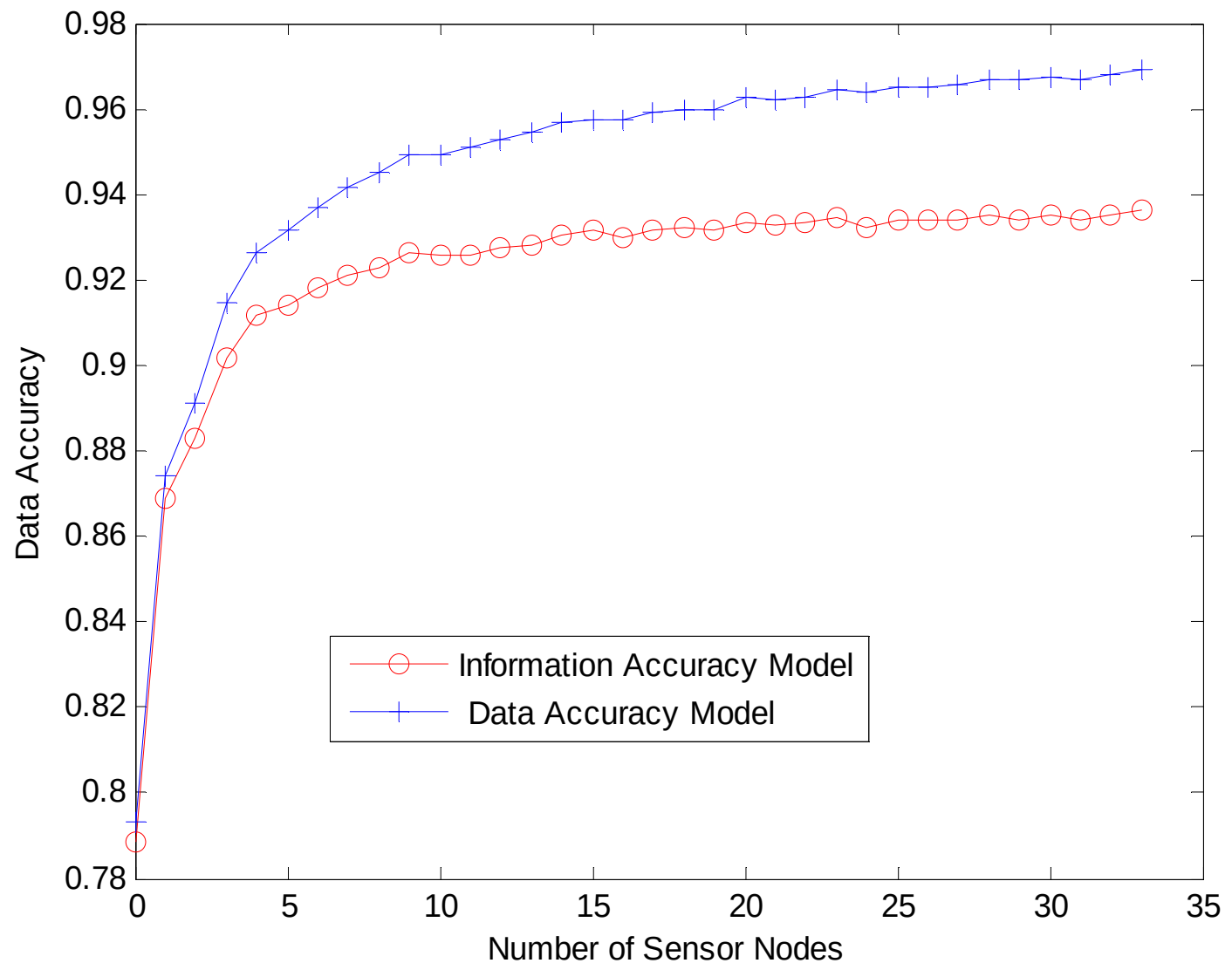

**Fig.3**. Number of sensor nodes versus data accuracy

In the third simulation setup, we have fixed number of sensor nodes ( $M = 4$ ), which forms a single cluster to sense and measure a single tracing point. We place four sensor nodes in a deployed circular cluster and set a tracing point at the central co-ordinate of deployed circular cluster as shown in Fig.4 (a). Since we have fixed number of sensor nodes, we vary the distance from $M$ number of sensor nodes from the tracing point $S$ in a deployed circular topology. As we increase the radius of the deployed circular cluster with same proportion from the tracing point $S$ as a centre, data accuracy decreases for the value $\theta_1$ =70 and $\theta_2$ =1. We compare our results from the results derived in literature [18] and conclude that as the radius of the deployed circular cluster increases with same proportion ,our Data accuracy model $D_A(M)$ always show better performance than Information accuracy model $I(M)$ with decreasing data accuracy as given in Fig.4(b).

In the fourth simulation set up, we have deployed thirty sensor nodes randomly in a sensor field of $100m \times 100m$ based sensor topology. Each sensor node $i$ has a data correlation radius $r = 23.5432$ of $cir(i)$ for an assumed threshold value $\tau = 0.5$. The neighboring sensor nodes $j$ which falls under the circular data correlation range $cir(i)$ around each node $i$ as a centre, the spatial observed data ( $S_i$ and $S_j$ ) are strongly correlated among them for which it is greater than equal to a threshold value( $\tau = 0.5$ ).Using this data correlation radius $r$ of each sensor node $i$ for $cir(i)$,we have developed a distributed clustering algorithm based on spatial data correlation among sensor nodes $i$ and $j$ as discuss in section-3. The sensor nodes form distributed non-overlapping clusters with irregular shape and size. Each distributed cluster can sense and measure a single tracing point located randomly with in the cluster. In a practical scenario, signal and noise variance of observed data changes with different location in the sensor field. For example the temperature variation changes from place to place in a tropical dense forest. Thus we adopt slightly different signal and noise variance of observed data for each distributed cluster in the sensor field. Once each distributed cluster can measure the observed data for tracing point $S$ , it calculates the data accuracy at the CH node for the respective cluster and finally transmits the most appropriate data to the sink node. Table.1 shows the comparison between Information accuracy model $I(M)$ and Data accuracy model $D_A(M)$ with respect to data accuracy for our proposed distributed clustering algorithm for each cluster. Each distributed cluster has its associate nodes along with a CH node where the data accuracy is performed. From Table.1, we can conclude that the result for our $D_A(M)$ gives more degree of data accuracy than $I(M)$ implemented in our clustering algorithm for each distributed cluster.

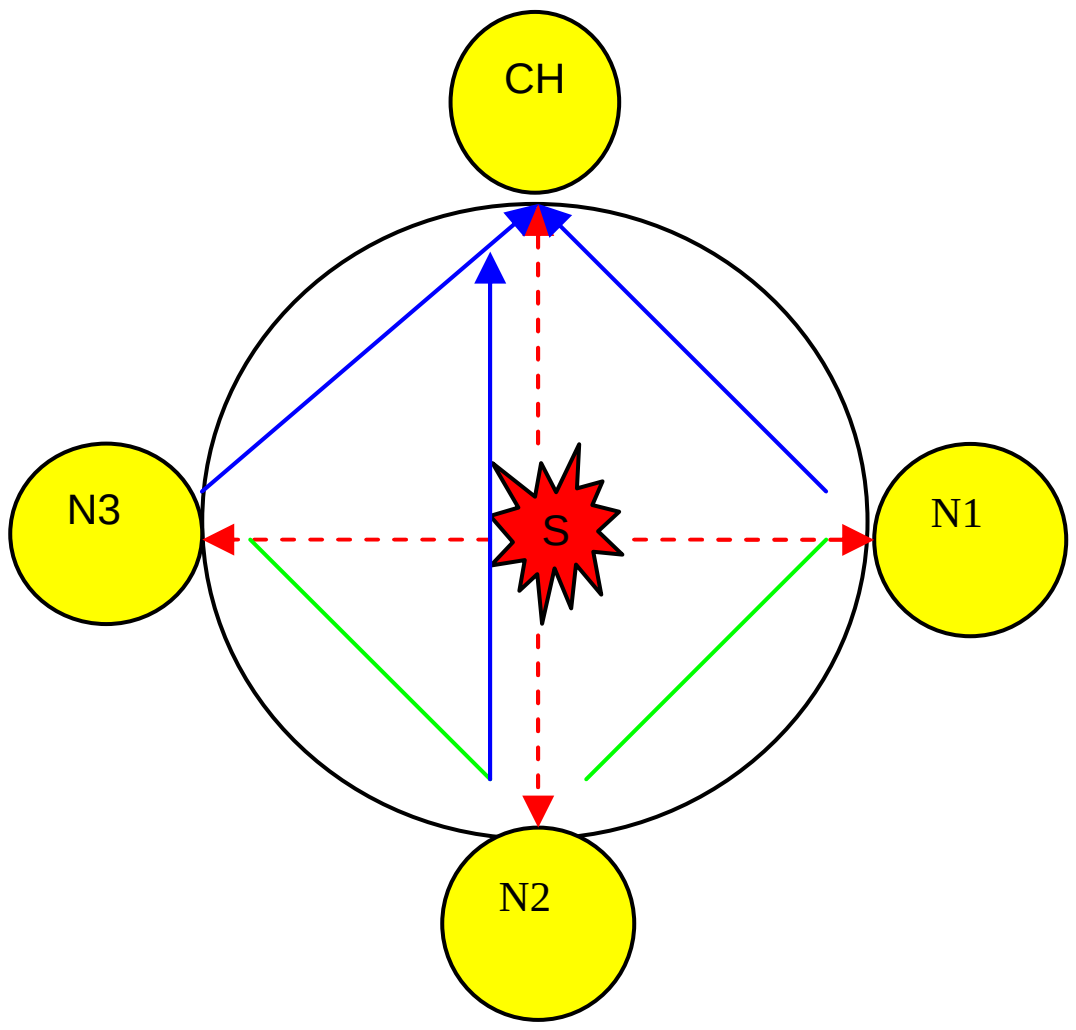

**Fig. 4(a)** Circular cluster topology with deployed nodes

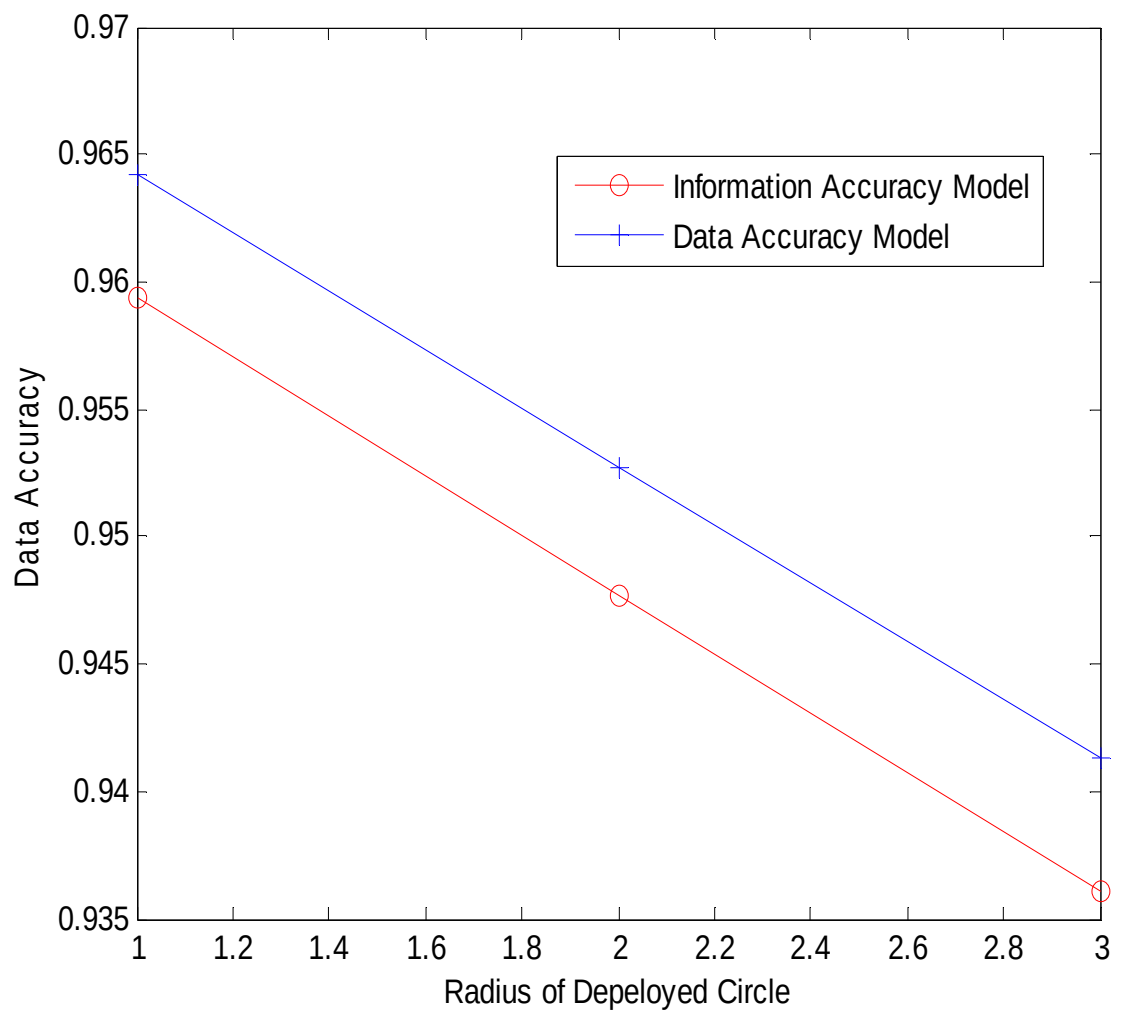

**Fig. 4(b)** Radius of circular cluster versus Data accuracy

| Cluster Number | Cluster head Node ID | Associated Nodes ID in Cluster | Information Accuracy $I(M)$ | Data Accuracy $D_A(M)$ |
|---|---|---|---|---|
| 1 | 29 | 1,3,5,9,14,23,24,25,30 | 0.8909 | 0.9748 |
| 2 | 7 | 4,12,19,20 | 0.8701 | 0.9462 |
| 3 | 26 | 6,8,15,16 | 0.8393 | 0.9541 |
| 4 | 13 | 2.10.21.22 | 0.8509 | 0.9660 |
| 5 | 11 | 18,27,28 | 0.9095 | 0.9701 |
| 6 | 17 | - | 0.9476 | 0.9476 |

**Table 1.** Data accuracy for each distributed cluster

## 6. Conclusions

We conclude in this paper that a non overlapping distributed clustering algorithm based upon data correlation among sensor nodes is proposed which reduces the data redundancy in the wireless sensor networks. We perform data accuracy for each distributed cluster at their respective CH node based on spatial correlation of data which shows that our propose Data accuracy model collects more accurate data and give better performance than Information accuracy model. Moreover our simulation results shows there exist an optimal cluster which is sufficient to perform approximately the same data accuracy level achieve by a cluster. In a cluster, the optimal cluster can perform the data accuracy at the CH node and rest of the sensor nodes goes in sleep mode. Thus it may reduce the communication overhead, energy consumption and increase the life time of distributed sensor networks.